\input{aipcheck}

\documentclass[
    ,final            
  ]
  {aipproc}

\layoutstyle{6x9}

\begin{document}

\vspace*{-15mm}
\begin{flushright}
SHEP--09--21\\
DFTT 61/2009\\
\end{flushright}
\vspace*{5mm}

\title{Unification of Gauge Couplings in the E$_6$SSM}


\author{P.~ Athron~}{
  address={Institut f\"ur Kern- und Teilchenphysik, TU Dresden, D-01062, Germany}}

\author{S.~ F.~ King~}{
  address={School of Physics and Astronomy, University of Southampton,
Southampton, SO17 1BJ, UK}
}

\author{R.~ Luo~}{
  address={Department of Physics and Astronomy, University of Glasgow,
Glasgow G12 8QQ, UK}
}

\author{D.~ J.~ Miller~}{
  address={Department of Physics and Astronomy, University of Glasgow,
Glasgow G12 8QQ, UK}
}

\author{S.~ Moretti~}{
  address={School of Physics and Astronomy, University of Southampton,
Southampton, SO17 1BJ, UK}
,altaddress={Dipartimento di Fisica Teorica, Universit\`a di Torino, 
Via Pietro Giuria 1, 10125 Torino, Italy}
}

\author{\qquad R.~ Nevzorov
\footnote{Based on a talk presented by R. Nevzorov at the SUSY'09 Conference,
Boston, USA, 5-10 June, 2009}
~\footnote{On leave of absence from the Theory Department, ITEP, Moscow, Russia}~
}{
  address={Department of Physics and Astronomy, University of Glasgow,
Glasgow G12 8QQ, UK}}

\begin{abstract}
We argue that in the two--loop approximation gauge coupling unification 
in the exceptional supersymmetric standard model (E$_6$SSM) can be achieved 
for any phenomenologically reasonable value of $\alpha_3(M_Z)$ consistent 
with the experimentally measured central value.
\end{abstract}

\maketitle


\section{Introduction}

The incorporation of weak and strong gauge interactions within Grand Unified Theories (GUTs) 
based on simple gauge groups such as $SU(5)$, $SO(10)$ or $E_6$ requires the unification of 
gauge couplings at some high energy scale $M_X$. On the other hand gauge coupling unification
makes also possible partial unification of gauge interactions with gravity in the framework
of superstring theories. At high energies the $E_6$ symmetry in the superstring inspired models 
can be broken to rank--5 subgroup $SU(3)_C\times SU(2)_W\times U(1)_Y\times U(1)'$ 
where $U(1)'=U(1)_{\chi} \cos\theta+U(1)_{\psi} \sin\theta$. Two anomaly-free $U(1)_{\psi}$ 
and $U(1)_{\chi}$ symmetries are defined by: $E_6\to SO(10)\times U(1)_{\psi},~
SO(10)\to SU(5)\times U(1)_{\chi}$. Here we concentrate on a particular $E_6$ inspired 
supersymmetric model with an extra $U(1)_{N}$ gauge symmetry that corresponds $\theta=\arctan\sqrt{15}$.
Only in this exceptional supersymmetric standard model (E$_6$SSM) \cite{1}-\cite{2} the right--handed 
neutrinos do not participate in gauge interactions. The extra $U(1)_{N}$ gauge symmetry survives 
to low energies and forbids a bilinear term $\mu H_d H_u$ in the superpotential of the 
considered model but allows the interaction $\lambda S H_d H_u$. At the electroweak (EW)
scale $S$ gets a non-zero vacuum expectation value (VEV), $\langle S \rangle=s/\sqrt{2}$, 
breaking $U(1)_N$ and an effective $\mu=\lambda s/\sqrt{2}$ term is automatically generated.

\section{The E$_6$SSM}

The E$_6$SSM is based on the $SU(3)_C\times SU(2)_W\times U(1)_Y \times U(1)_N$ 
gauge group which is a subgroup of $E_6$. To ensure anomaly cancellation the particle
content of the E$_6$SSM is extended to include three complete {\bf $27$} representations 
of $E_6$ \cite{1}-\cite{2}. Each {\bf $27_i$} multiplet contains a SM family of quarks and 
leptons, right--handed neutrino $N^c_i$, SM singlet field $S_i$ which carries a non--zero 
$U(1)_{N}$ charge, a pair of $SU(2)_W$--doublets $H_{1i}$ and $H_{2i}$ with the quantum numbers 
of Higgs doublets and a pair of colour triplets of exotic quarks $\overline{D}_i$ and $D_i$
which can be either diquarks (Model I) or leptoquarks (Model II) \cite{1}-\cite{2}. 
$H_{1i}$ and $H_{2i}$ form either Higgs or inert Higgs multiplets. We also require a 
further pair $L'$ and $\overline{L}'$ from incomplete extra $27'$ and $\overline{27'}$ 
representations to survive to low energies to ensure gauge coupling unification in 
the one--loop approximation. The presence of a $Z'$ boson and exotic quarks predicted by 
the E$_6$SSM provides spectacular new physics signals at the LHC which were discussed in 
\cite{1}-\cite{30}. In the two--loop approximation the lightest Higgs boson mass in the 
E$_6$SSM does not exceed $150-155\,\mbox{GeV}$ \cite{1}-\cite{2}. Recently the particle 
spectrum within the constrained version of the E$_6$SSM was studied \cite{4}-\cite{50}.

Since right--handed neutrinos have zero charges in the considered model they are expected 
to gain Majorana masses at some intermediate scale. The heavy Majorana neutrinos 
may decay into final states with lepton number $L=\pm 1$, creating a lepton asymmetry 
in the early Universe. Due to the presence of exotic particles the substantial values of the 
CP asymmetries in the E$_6$SSM can be induced even for a relatively small mass of 
the lightest right--handed neutrino ($M_1 \sim 10^6\,\mbox{GeV}$) so that the successful 
thermal leptogenesis may be achieved without encountering gravitino problem \cite{6}. 

The superpotential in $E_6$ inspired models involves a lot of new Yukawa couplings that induce 
non--diagonal flavour transitions. To avoid a flavour changing neutral current (FCNC) problem 
an approximate $Z^{H}_2$ symmetry is postulated in the E$_6$SSM. Under this symmetry all 
superfields except $H_d\equiv H_{13}$, $H_u\equiv H_{23}$ and $S\equiv S_3$
are odd. The $Z^{H}_2$ symmetry reduces the structure of the Yukawa interactions to:
\begin{equation}
\begin{array}{rcl}
W_{\rm ESSM}&\simeq & \lambda_i S(H_{1i}H_{2i})+\kappa_i
S(D_i\overline{D}_i)+f_{\alpha\beta}S_{\alpha}(H_d
H_{2\beta})+ \tilde{f}_{\alpha\beta}S_{\alpha}(H_{1\beta}H_u)+\\
&&+\mu'(L'\overline{L}')+h^{E}_{4j} (H_d L') e^c_j +W_{\rm{MSSM}}(\mu=0),
\end{array}
\label{1}
\end{equation}
where $\alpha,\beta=1,2$ and $i=1,2,3$\,. Here we assume that all right--handed neutrinos are 
heavy. The $SU(2)_W$ doublets $H_u$ and $H_d$ play the role of Higgs fields. Therefore it is 
natural to assume that only $S$, $H_u$ and $H_d$ acquire VEVs. The VEV of the field $S$ breaks 
$U(1)_N$ symmetry inducing effective $\mu$ term as well as the masses of exotic fermions and 
$Z'$ boson. To guarantee that only $H_u$, $H_d$ and $S$ develop VEVs in the E$_6$SSM a certain 
hierarchy between the Yukawa couplings is imposed, i.e. 
$\kappa_i\sim\lambda_i\gg f_{\alpha\beta}, \tilde{f}_{\alpha\beta}, h^{E}_{4j}$.

\section{RG flow of gauge couplings}

At high energies the two--loop renormalisation group (RG) flow of gauge couplings in the E$_6$SSM 
can be parametrised as
\begin{equation}
\frac{1}{\alpha_i(t)}=\frac{1}{\alpha_i(M_Z)}-\frac{b_i}{2\pi} t-\frac{C_i}{12\pi}-\Theta_i(t)
+\frac{b_i-b_i^{SM}}{2\pi}\ln\frac{T_i}{M_Z}\,, \qquad 
T_i=\prod_{k=1}^N\biggl(m_k\biggr)^{\frac{\Delta b^k_i}{b_i-b_i^{SM}}}\,,
\label{2}
\end{equation}
where index $i$ runs from $1$ to $3$ and corresponds to $U(1)_Y$, $SU(2)_W$ and $SU(3)_C$ 
interactions, $b_i$ and $b_i^{SM}$ are the coefficients of the one--loop beta functions in the 
E$_6$SSM and SM, $t=\ln\left(\mu/M_Z\right)$, $\mu$ is a renormalisation scale, $C_1=0$, 
$C_2=2$, $C_3=3$, $\Theta_i(t)$ are two--loop corrections and $T_i$ are effective
threshold scales. In Eq.~(\ref{2}) $m_k$ and $\Delta b_i^k $ are masses of new particles and 
their contributions to the beta functions. In the E$_6$SSM the one--loop beta functions are 
$b_1=48/5$, $b_2=4$, $b_3=0$. The two--loop contributions to the beta functions of gauge 
couplings in the E$_6$SSM were calculated in \cite{8}.

Using the approximate solution of the RG equations (RGEs) in Eq.~(\ref{2}) one 
can find the value of $\alpha_3(M_Z)$ for which exact gauge coupling unification can be
achieved:
\begin{equation}
\frac{1}{\alpha_3(M_Z)}=\frac{1}{b_1-b_2}\biggl[\frac{b_1-b_3}{\alpha_2(M_Z)}-
\frac{b_2-b_3}{\alpha_1(M_Z)}\biggr]-\frac{1}{28\pi}+\Theta_s-\Delta_s\,,\quad 
\Delta_s=-\frac{19}{28\pi}\ln\frac{\tilde{M}_{S}}{M_Z}\,,
\label{31}
\end{equation}
\begin{equation}
\Theta_s=\biggl(\frac{b_2-b_3}{b_1-b_2}\Theta_1-\frac{b_1-b_3}{b_1-b_2}\Theta_2+\Theta_3\biggr)\,,
\qquad \Theta_i=\Theta_i(M_X),
\label{32}
\end{equation}
where the effective threshold scale $\tilde{M}_S$ can be expressed in terms 
of the MSSM one $M_S$:
\begin{equation}
\tilde{M}_S=\frac{T_2^{172/19}}{T_1^{55/19} T_3^{98/19}}=
M_S\cdot\frac{m_{H_{\alpha}}^{12/19}\mu_{\tilde{H}_{\alpha}}^{24/19}\mu^{'18/19}}
{m_{\tilde{D}_i}^{18/19}\mu_{D_i}^{36/19}}\simeq \mu'\cdot \biggl(\frac{M_{weak}}{M_{colour}}\biggr)^{4.5}\,.
\label{5}
\end{equation}
In Eq.~(\ref{5}) $\mu_{D_i}$ and $m_{\tilde{D}_i}$ are the masses of exotic quarks and their superpartners, 
$m_{H_{\alpha}}$ and $\mu_{\tilde{H}_{\alpha}}$ are the masses of inert Higgs and inert Higgsino fields, 
while the scalar and fermion components of $L'$ and $\overline{L}'$ are degenerate around $\mu'$.

Because the unification of gauge couplings is determined by only one effective threshold
scale $\tilde{M}_S$ one can simplify the analysis assuming that $T_1=T_2=T_3=\tilde{M}_S$.
In our numerical study we parametrise the $\tilde{M}_S$ in terms of two scales. One of these scales
is associated with the SUSY breaking scale $M_S$ while another one is set by the masses of $Z'$ and
exotic fermions and bosons which we assume to be degenerate around $M_{Z'}$ for simplicity.
Thus we use the SM beta functions to describe the running of $\alpha_i(t)$ between $M_Z$ and
$M_S$, then we apply the two--loop RGEs of the MSSM to compute the flow of $\alpha_i(t)$ from $M_S$
to $M_{Z'}$ and the two--loop RGEs of the E$_6$SSM to calculate the evolution of $\alpha_i(t)$
between $M_{Z'}$ and $M_X$ which is equal to $3.5\cdot 10^{16}\,\mbox{GeV}$ in the case
of the E$_6$SSM. Since $M_S\simeq \mu/6$ we choose $M_S\simeq 250\,\mbox{GeV}$.
We also fix $M_{Z'}=1.5\,\mbox{TeV}$.

\begin{figure}
\begin{tabular}{cc}
\includegraphics[height=.2\textheight]{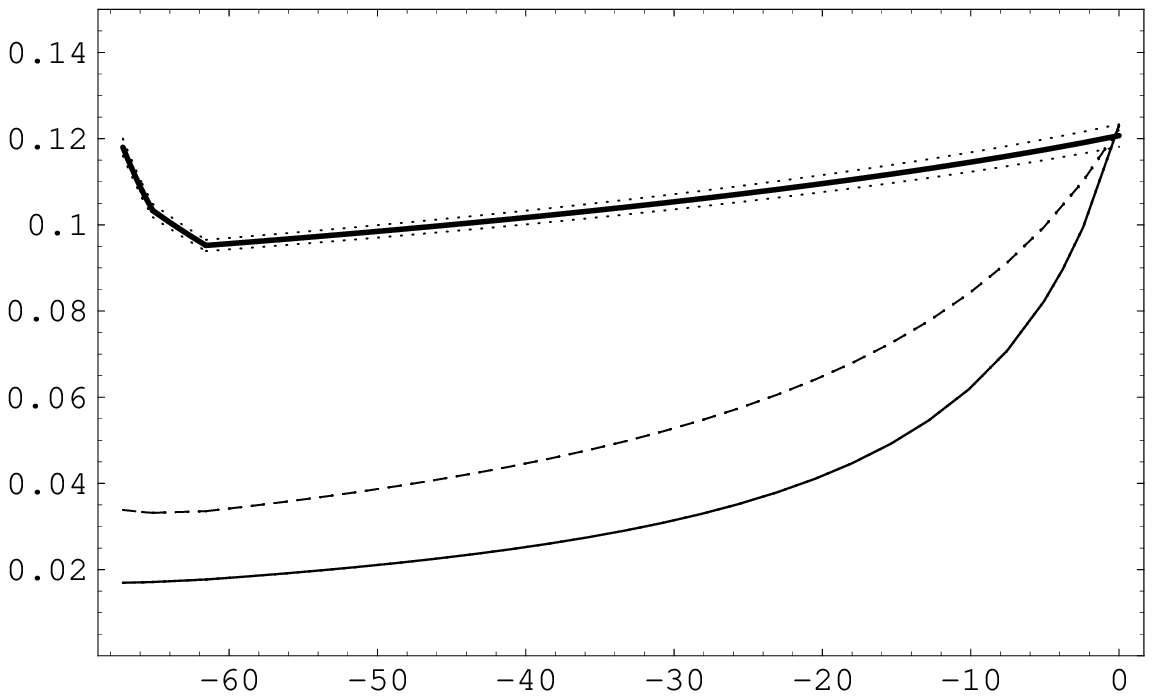} &
\includegraphics[height=.2\textheight]{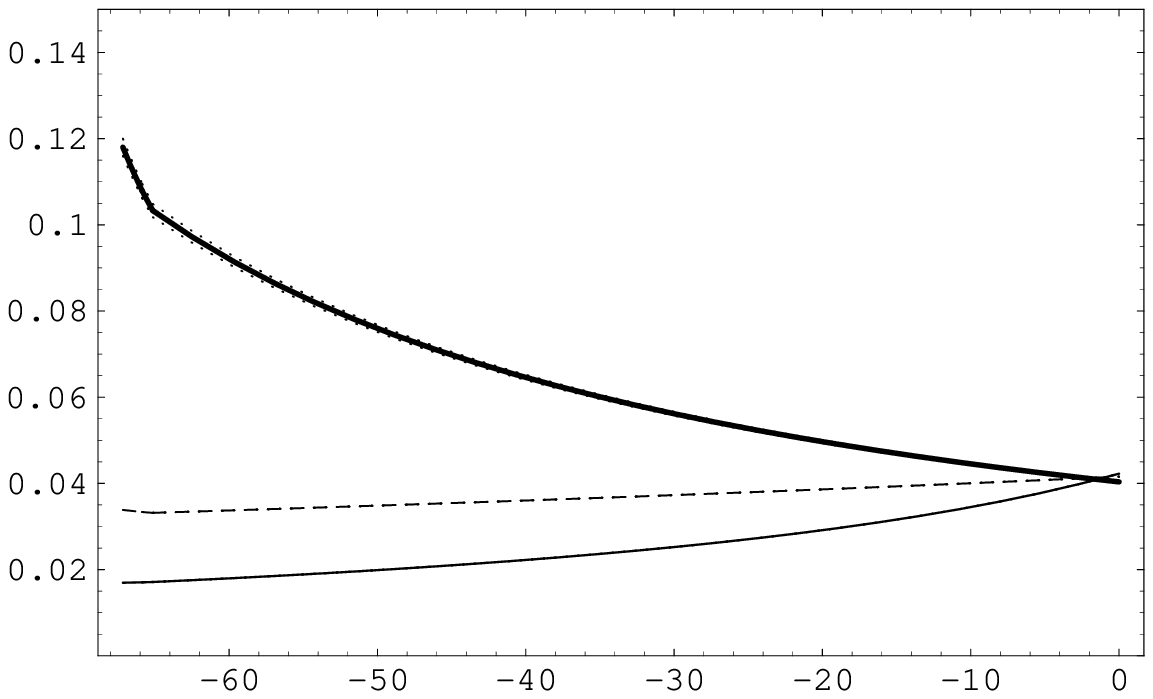} 
\caption{Two--loop RG flow of gauge couplings from EW to GUT scale $M_X$ in the E$_6$SSM (left) 
and MSSM (right). Thick, dashed and solid lines correspond to the running of $SU(3)_C$, $SU(2)_W$ 
and $U(1)_Y$ couplings respectively. We used  $\tan\beta=10$, $M_S=250\,\mbox{GeV}$, 
$M_{Z'}=1.5\,\mbox{TeV}$, $\alpha_s(M_Z)=0.118$, $\alpha(M_Z)=1/127.9$ and $\sin^2\theta_W=0.231$. 
The dotted lines represent the uncertainty in $\alpha_i(t)$ caused by the variation of the strong 
gauge coupling from 0.116 to 0.120 at the EW scale.}
\end{tabular}
\end{figure}

The results of our numerical studies of gauge coupling unification are summarised in Fig.~1.
where the RG flow of gauge couplings in the E$_6$SSM and MSSM are shown. 
Due to the presence of exotic matter the E$_6$SSM gauge couplings are considerably larger at 
high energies than in the MSSM. Dotted lines in Fig.~1 represent the changes of the evolution 
of $\alpha_i(t)$ induced by the variations of $\alpha_3(M_Z)$ within $1\,\sigma$ around its 
average value. In the E$_6$SSM the two--loop effects lead to the mild growth of $\alpha_3(t)$
as renormalisation scale increases. As a consequence at high energies the uncertainty in $\alpha_3(t)$ 
caused by the variations of $\alpha_3(M_Z)$ is much bigger in the E$_6$SSM than in the MSSM. 
The relatively large uncertainty in $\alpha_3(M_X)$ allows one to achieve exact unification of 
$\alpha_i(t)$ even within $1\,\sigma $ deviation of $\alpha_3(M_Z)$ from its average value.

It is worth noticing here that two--loop corrections to $\alpha_i(t)$ are large in the 
E$_6$SSM and could spoil the gauge coupling unification. Nevertheless the value of $\alpha_3(M_Z)$ 
that results in the exact gauge coupling unification in the E$_6$SSM is quite close to the 
experimentally measured central value. Without the inclusion of threshold effects Eq.~(\ref{31}) 
gives $\alpha_3(M_Z)\simeq 0.121$ which is very close to the one--loop prediction for $\alpha_3(M_Z)=0.118$.
The small difference between one--loop and two--loop predictions for $\alpha_3(M_Z)$
is caused by the remarkable cancellation of different two--loop corrections in Eq.~(\ref{32}).

As in the MSSM the inclusion of threshold effects lowers the prediction for the value of the 
strong gauge coupling at the EW scale. From Eq.~(\ref{5}) it is obvious the effective threshold 
scale in the E$_6$SSM is set by $\mu'$. The term $\mu'L'\overline{L}'$ in the superpotential is 
not involved in the process of EW symmetry breaking and is not suppressed by the $E_6$ symmetry. 
Therefore, although the effective threshold scale $\tilde{M}_S$ may be considerably less than 
$\mu'$, the corresponding mass parameter can be always chosen so that $\tilde{M}_S$ lies in a 
few hundred GeV range that allows to get the exact unification of gauge couplings for any value 
of $\alpha_3(M_Z)$ which is in agreement with current data. 

\begin{theacknowledgments}
RN acknowledges support from the SHEFC grant HR03020 SUPA 36878.
SM is financially supported in part by the scheme `Visiting Professor - Azione D - 
Atto Integrativo tra la Regione Piemonte e gli Atenei Piemontesi.
\end{theacknowledgments}

\end{document}